\documentclass[useAMS,usenatbib]{mn2e}

\usepackage{graphicx}

\newcommand{\gtapprox}{\raisebox{-0.5ex}{$\,\stackrel{>}{\scriptstyle
\sim}\,$}}
\newcommand{\ltapprox}{\raisebox{-0.5ex}{$\,\stackrel{<}{\scriptstyle
\sim}\,$}}
\newcommand{\ri}{r_{\rm i}}
\newcommand{\rg}{r_{\rm g}}
\newcommand{\UB}{U_{\scriptscriptstyle \rm B}}
\newcommand{\epsIC}{\varepsilon^{\scriptscriptstyle \rm IC}}
\newcommand{\epsSSC}{\varepsilon^{\scriptscriptstyle \rm SSC}}
\newcommand{\epssyn}{\varepsilon^{\rm syn}}
\newcommand{\fss}{f_{\rm ss}}
\newcommand{\fj}{f_{\rm j}}
\newcommand{\Fd}{F_{\rm d}}
\newcommand{\LEdd}{L_{\rm Edd}}
\newcommand{\Ld}{L_{\rm d}}

\newcommand{\Lx}{L_{0.5-8 \,{\rm keV}}}
\newcommand{\betaj}{\beta_{\rm j}}
\newcommand{\Gammaj}{\Gamma_{\rm j}}
\newcommand{\thetai}{\theta_{\rm i}}
\newcommand{\gammamin}{\gamma_{\rm min}}
\newcommand{\gammamax}{\gamma_{\rm max}}
\newcommand{\phij}{\phi_{\rm j}}
\newcommand{\rj}{r_{\rm j}}

\newcommand{\Pa}{P_{\rm a}}

\newcommand{\feq}{f_{\rm eq}}
\newcommand{\Ue}{U_{\rm e}}
\newcommand{\Ne}{N_{\rm e}}
\newcommand{\Ke}{K_{\rm e}}
\newcommand{\Pj}{P_{\rm j}}
\newcommand{\Pbke}{P_{\rm bke}}
\newcommand{\Pe}{P_{\rm e}}
\newcommand{\PB}{P_{\scriptscriptstyle \rm B}}

\newcommand{\sigmaT}{\sigma_{\scriptscriptstyle \rm T}}
\newcommand{\Mdota}{\dot M_{\rm a}}
\newcommand{\dr}{{\rm d}r}

\newcommand{\Mbh}{M_{\rm bh}}
\newcommand{\RX}{R_{\scriptscriptstyle \rm X}}
\newcommand{\LX}{L_{\scriptscriptstyle \rm X}}
\newcommand{\LR}{L_{\scriptscriptstyle \rm R}}

\title[Spectral properties of ULX Models]{Radio and X-ray Properties of
Relativistic Beaming Models for Ultra-Luminous X-ray Sources}
\author[Freeland et al.]{M. Freeland$^1$\thanks{E-mail:
mcf35@cam.ac.uk, z.kuncic@physics.usyd.edu.au, rs1@mssl.ucl.ac.uk, geoff@mso.anu.edu.au},
Z. Kuncic$^1$\footnotemark[1],
R. Soria$^{2,3}$\footnotemark[1],
and G. V. Bicknell$^4$\footnotemark[1]
\\
$^1$School of Physics, University of Sydney, NSW 2006, Australia \\
$^2$Harvard-Smithsonian Centre for Astrophysics, Cambridge, MA 02138, USA\\
$^3$Mullard Space Science Laboratory, University College London, Holmbury St Mary,
Dorking, RH5 6NT, UK\\
$^4$Research School of Astronomy \& Astrophysics, Australian National University, ACT 2611, Australia
}
\begin{document}

\date{Accepted 1988 December 15. Received 1988 December 14; in original form 1988 October 11}

\pagerange{\pageref{firstpage}--\pageref{lastpage}} \pubyear{2002}

\maketitle

\label{firstpage}

\begin{abstract}
We calculate the broadband radio--X-ray spectra predicted by microblazar and
microquasar models for Ultra-Luminous X-ray sources (ULXs),
exploring the possibility that their dominant power-law component is produced
by a relativistic jet, even at near-Eddington mass accretion rates.
We do this by first constructing a generalized disk--jet theoretical framework in
which some fraction of the total accretion power $\Pa$ is efficiently removed
from the accretion disk by a magnetic torque responsible for jet formation.
Thus, for different black hole masses, mass accretion rates
and magnetic coupling strength, we self-consistently calculate the relative importance
of the modified disk spectrum, as well as the overall jet emission due to synchrotron
and Compton processes.
In general, transferring accretion power to a jet makes the disk fainter and cooler
than a standard disk at the same mass accretion rate;
this may explain why the soft spectral component appears less prominent than the
dominant power-law component in  most bright ULXs.
We show that the apparent X-ray luminosity and spectrum predicted by the microquasar
model are consistent with the observed properties of most ULXs.
We predict that the radio synchrotron jet emission is too faint to be detected
at the typical threshold of radio surveys to date.
This is consistent with the high rate of non-detections over detections in
radio counterpart searches.
Conversely, we conclude that the observed radio emission found associated with
a few ULXs cannot be due to beamed synchrotron emission from a relativistic jet.
\end{abstract}

\begin{keywords}
accretion, accretion disks -- black hole physics -- X-rays: binaries -- relativistic jets.
\end{keywords}

\section{Introduction}

Ultra-Luminous X-ray sources (ULXs) are among the most intriguing sources to have
been discovered by X-ray satellites.
They are defined as point-like X-ray sources which, if located within
the galaxy they are associated with on the sky, are off-nuclear and have X-ray
luminosities $\Lx \gtapprox \, 2 \times 10^{39}\, {\rm erg \, s}^{-1}$
in the $0.5 - 8.0\,$keV bandpass \citep*{Irwin04}.
This exceeds the theoretical Eddington limit for a neutron star or stellar-mass black hole,
$L_{\rm Edd} = 4\pi G\Mbh \mu m c/\sigmaT$, where $\Mbh$ is the black hole mass and $\mu$
is the mean molecular weight.
It also exceeds the peak in observed X-ray luminosities of known Galactic X-ray binaries
(XRBs).
Spectroscopic studies have revealed that some ULXs can be identified with
supernova remnants or background AGN \citep[see e.g.][and references therein]{GutLopez05}.
Of the remaining unidentified sources, however, rapid X-ray variability
\citep*[e.g.][]{StroMush03,SoriaMotch04,Krauss05}
and spectral state transitions
\citep*{Kubota01,Kubota02,LaParola01,MizKubMak01,Strickland01}
strongly suggest that we are dealing
with a population of close interacting binary systems involving accretion onto a black hole.

Perhaps the most contentious issue surrounding ULXs is the mass of the black hole, $\Mbh$,
which determines $L_{\rm Edd}$.
There are at least three possible explanations for why many ULXs are apparently
$\sim 20-30$ times  more luminous than typical Galactic black hole XRBs:
1. The emission is Doppler-boosted by a relativistic jet pointing towards us
(microblazar scenario: e.g. \citealt*{FabMesch01,Kording02}).
In this case, $\Mbh$ need not be more than $\sim \, \mbox{a few} \, M_\odot$, similar
to typical stellar-mass black holes in the Galaxy.
2. The brightness enhancement is due to a combination of various factors: mild
beaming, $\Mbh$ more massive than typical Galactic sources, and super-Eddington accretion.
A factor of $\sim 3$ for each of these terms is physically plausible and would
suffice for most ULXs \citep[e.g.][]{KingDehnen05}.
In particular, mild geometric beaming could result from a sub-relativistic,
radiatively-driven disk outflow (\citealt{King01,King04}).
Alternatively, it could be the result of a relativistic jet pointing slightly away
from us (microquasar scenario).
3. The X-ray emission is isotropic and Eddington-limited; the high luminosity arises
from a more massive black hole (an intermediate-mass black hole, IMBH), with
$\Mbh \sim 10^{2-3} M_\odot$  \citep*{ColMush99,Makishima00,Miller03}.
In this case the X-ray source is truly ultra-luminous, and hence it requires an
accretion rate higher than that for a beamed scenario.
If ULXs were proven to be powered by an accreting IMBH, the implications would be
far-reaching, impacting on many other fields, from star-formation to cosmology
(see \citealt{Miller04} for a review).

In principle, phase-resolved optical spectroscopic and photometric studies are the
most direct way to determine the binary system parameters (viewing angle,
orbital period, radial velocity curve and hence, mass function --
see e.g. \citealt{Charles98}) and constrain the nature of ULXs.
This is how Galactic black holes were first identified.
However, optical counterpart searches and phase-resolved studies have proven
exceedingly difficult for ULXs because even the nearest ones are located at
distances of a few Mpc.
Moreover, the brightest ULXs tend to be preferentially located in crowded star
forming regions  \citep{Irwin04}, making unambiguous optical identifications
often impossible, even within the \textit{Chandra} error circle.
When optical counterparts are found, they are generally consistent with O or B0
stars \citep{Liu02,Liu04,Kaaret04,Zampieri04}.
However, this could be a selection effect, since smaller donors (perhaps
lower-mass stars evolving through their giant phase) would be too faint to be
detected.
Hence, it is important to identify the discriminating signatures of ULXs
in other wavebands.

Radio observations may provide a more effective means of differentiating
the isotropic IMBH and beaming models.
By analogy with AGN, we can expect a ULX microblazar to produce strong
synchrotron radio emission from a compact, unresolved core, while more extended,
and possibly elongated and resolved radio structure may arise from a ULX microquasar,
seen at larger inclination.
To date, radio counterparts have been detected for only a few ULXs:
two in M82 \citep*{KronSram85,KordColFalck05};
one in NGC\,5408 \citep*{Kaaret03,Soria06a};
one in Holmberg\,II \citep*{TonWest95,MillMushNeff05};
one in NGC\,7424 \citep*{Soria06b}; and one in NGC\,6946
(van Dyk et al. 1994; Blair, Fesen \& Schlegel, 2001; Swartz et al. 2006, in prep.).
It is still unclear whether the emission in those radio sources is due to an
underlying supernova remnant or is directly produced by ULX jet activity, either
from the core or the lobes.
Hence, it is important to determine the radio signatures of different ULX models.

In the X-ray band, ULX spectra are generally dominated by a power-law, particularly
above 1~keV.
A cool thermal component is present in some sources,
but contributes at most $10 - 20\,$\% of the flux \citep*[e.g.][]{Stobbart06}.
A power-law component is also dominant in the X-ray spectra of Galactic black holes
in two of the canonical spectral states \citep{McClinRem06}: the low/hard state,
when the photon index is $\Gamma \sim 1.5 - 2$ and the X-ray luminosity
is $\Lx \ltapprox 10^{-2} L_{\rm Edd}$; and the very high (or steep power-law)
state, when $\Gamma \gtapprox 2.5$ and $\Lx \sim L_{\rm Edd}$.
In the low/hard state, the accretion disk is thought to be truncated far from the
innermost stable circular orbit (ISCO), and replaced by a radiatively inefficient
flow with a steady jet \citep[see e.g.][]{Esin97}.
The very high state, on the other hand, is characterized by a high radiative
efficiency, quasi-periodic X-ray variability, and absence of a steady jet,
although flares or sporadic radio ejections are associated with this state.
It is not clear whether ULXs can be classified into either one of these power-law
dominated XRB states; their photon index is somewhat  intermediate, with
$\Gamma \approx 1.5 - 2.5$ (e.g. \citealt{Swartz04,Stobbart06,Winter06}).
If they are in a classical low/hard state, their apparent X-ray luminosities
suggest masses $\gtapprox \mbox{a few} \times 10^3 M_\odot$.
What seems clear is that ULXs are not in a high/soft state, when the X-ray spectrum
is dominated by soft disk blackbody emission and the power-law is
weaker.

Regardless of which spectral state ULXs belong to,
the power-law component could be produced either by thermal Comptonization
in a hot, diffuse corona, or by nonthermal synchrotron and Compton processes
in a relativsitic jet, or possibly a combination of both
\citep*[see][for a discussion]{Markoff03}.
Although it is widely accepted that relativistic jets are responsible for extragalactic
radio sources, analogous jet models for Galactic XRBs have only been considered
relatively recently \citep*[see][for a review]{Fender05}.
Existing disk-jet models for binaries, however, either neglect the radio synchrotron
properties altogether \citep*[e.g.][]{GeorgAharKirk02}
or are only appropriate for the low/hard state 
\citep*[e.g.][]{MarkFalckFend01,Markoff03,FendBellGall04}.
It is important to explore the possibility of a jet coexisting with 
(and coupled to) a bright disk extending all the way down to the
ISCO\footnote{Recent X-ray/radio studies of FRII radio galaxies and quasars
(e.g. \citealt{PunslyTingay05,BallFab05}) provide examples
of powerful nuclear jets co-existing with a high luminosity spectral state
($\LX \sim L_{\rm Edd}$).}.

In this paper, we investigate both the radio and X-ray properties of relativistic
beaming models for ULXs using an accreting black hole framework generalized to include
coupled disk-jet spectral components.
Specifically, the radio and hard X-ray emission are produced in a relativistic jet
that is magnetically coupled to an accretion disk (which can contribute 
to the soft X-ray emission).
The jet drains a substantial fraction of the total accretion power from the disk
and thus modifies the multi-colour disk spectrum.
In order to calculate the spectral energy distribution, we explicitly model the mechanism
which magnetically couples the disk and jet. 
Thus, our generalized model calculates both the jet and disk spectra
\textit{self-consistently} for different values of the black hole mass $\Mbh$,
mass accretion rate $\Mdota$, and coupling strength.
This leads to two unique features of the model that have important implications not just
for ULXs, but for all accreting systems:
1. The nett accretion power can exceed the Eddington limit because the jet efficiently
channels energy from the disk, which remains sub-Eddington; and
2. The presence of a jet can be inferred indirectly from its effects on the disk,
which consequently emits a spectrum that can be considerably different from that
predicted by the standard Shakura-Sunyaev \citep{SS73} disk in the EUV/soft-X-ray
bandpass for XRBs.
Our coupled disk-jet theory
is outlined in Sec.~\ref{s:theory}; results for beamed ULX models are 
presented in Sec.~\ref{s:results}; and a discussion and conclusions are given in
Secs.~\ref{s:discussion} and \ref{s:conclusions}, respectively.

\section{Theoretical Basis}
\label{s:theory}

The model we construct here is based on the generalized accretion disk theory of
\cite{KunBick04}, which gives self-consistent solutions for the radiative flux of
a turbulent, magnetized accretion disk modified by an outflow that could be either
mass-flux or Poynting-flux dominated.
A mass-flux dominated outflow requires $\Mdota$ to vary with radius $r$ in the disk.
We  restrict ourselves here to modelling a Poynting-flux dominated outflow.
The solutions also take into account the possibility of a non-zero magnetic torque
acting on the inner boundary, although we do not consider this effect here.
We calculate an outflow-modified multi-colour-disk (OMMCD) spectrum, assuming
each annulus is locally emitting a blackbody.
A Newtonian potential is used.

The fraction of accretion power removed from the disk by the magnetic torque acting
on the surface is channelled into the jet and partitioned into kinetic energy
(bulk and random), magnetic field, and radiative energy components.
We calculate the jet spectrum due to synchrotron, synchrotron self-Comptonization (SSC)
and inverse Compton (IC) scattering of the disk photons by the energetic electrons
in the jet.
We test the micro-blazar/quasar models for ULXs by comparing the theoretical spectra for
different values of $\Mbh$, $\Mdota$, jet power $\Pj$ and inclination angle $\thetai$
measured relative to the jet axis.

\subsection{Modified disk emission}
\label{s:ommcd}

Assuming a zero nett torque at the inner disk boundary $\ri$ and no radial variations in
the mass accretion rate $\Mdota$ due to a mass-loaded wind, the radiative flux of a
magnetized accretion disk with a non-zero magnetic torque acting on the disk surface is
\citep[][2006, in preparation]{KunBick04}
\begin{equation}
 \Fd (r) = \frac{3G \Mbh \Mdota}{8\pi r^3} [ \fss (r) - \fj (r) ] \qquad ,
\end{equation}
where $\fss (r)$ is the small-$r$ correction term in the standard Shakura-Sunyaev (SS) disk
theory, given by  \citep{SS73}
\begin{equation}
\fss = 1 - \left( \frac{r}{\ri} \right)^{-1/2} \qquad .
\end{equation}
The term $\fj(r)$ is the jet correction factor due to a non-zero magnetic torque on the disk
surface:
\begin{equation}
\fj (r) = \frac{1}{\Mdota \Omega r^2} \int_{\ri}^{r} 4\pi r^2 \frac{B_\phi^+ B_z^+}{4\pi}\, \dr
\qquad ,
\label{e:fj}
\end{equation}
where $B_\phi^+$ and  $B_z^+$ are the azimuthal and vertical components of the local
magnetic field, evaluated at the disk surface, and $\Omega = (G\Mbh / r^3)^{1/2}$ is
the Keplerian orbital velocity.

We assume a power-law radial profile for the magnetic stress, \textit{viz.}
\begin{equation}
\frac{B_\phi^+ (r) B_{z}^{+}(r)}{4\pi} = \frac{B_{\phi}^{+}(\ri) B_{z}^{+}(\ri)}{4\pi}
\left( \frac{r}{\ri} \right)^{-q}
\label{e:q}
\end{equation}
and we require $q \gtapprox 2$ so that the total work done against the disk surface by the
magnetic torque remains finite.
The normalization of the magnetic stress is determined from global energy conservation,
which requires
\begin{equation}
\Ld = \Pa - \Pj \qquad ,
\end{equation}
where 
\begin{equation}
\Ld = 2 \int_{\ri}^{\infty} 2 \pi r \Fd (r) \, \dr
\end{equation}
is the disk radiative power,
\begin{equation}
\Pa = \int_{\ri}^\infty 4 \pi r \frac{3G\Mbh \Mdota}{8\pi r^3} \fss (r) \, \dr =
\frac{1}{2} \frac{G \Mbh \Mdota}{\ri}
\label{e:Pa}
\end{equation}
is the total accretion power, and 
\begin{equation}
\Pj = \int_{\ri}^\infty 4 \pi r \frac{3G \Mbh \Mdota}{8\pi r^3} \fj (r) \, \dr
\label{e:Pj}
\end{equation}
is the total jet power.
Thus, the magnitude of the magnetic stress at $\ri$ can be related to the fraction
$\Pj / \Pa$ of total accretion power channelled into the jet, and substitution into
(\ref{e:fj}) then yields
\begin{equation}
\fj (r) = \frac{1}{2} \left( \frac{\Pj}{\Pa} \right)
\left( \frac{q-\frac{3}{2}}{q-3} \right) \left( \frac{r}{\ri} \right)^{-1/2}
\left[ 1 - \left( \frac{r}{\ri} \right)^{(3-q)} \right]
\label{e:fjr}
\end{equation}
For the specific case $q=2.5$, this simplifies to $\fj = (\Pj / \Pa)\fss$ and hence,
the disk flux is just $\Fd (r) = (3G \Mbh \Mdota /8\pi r^3)\fss (1 - \Pj/\Pa)$,
which has the same radial dependence as the SS disk.

The total disk spectrum is calculated assuming local blackbody emission, $B_\nu$,
at each $r$:
\begin{equation}
L_{\rm d,\nu} = \int_{\ri}^\infty 4 \pi^2 r B_\nu [T(r)] \, \dr \qquad ,
\end{equation}
where $T(r) = \left[ \Fd (r) /\sigma \right]^{1/4}$ is the effective disk temperature at
each radius and $\sigma$ is the Stefan--Boltzmann constant.
For a disk inclined at an angle $\thetai$ with respect to an observer's line-of-sight,
the apparent luminosity spectrum is
$L_{\rm d, \nu}^{\rm obs} =\frac{1}{2} \cos \thetai L_{\rm d, \nu}$.
The effective temperature may not be directly observable; what is typically fitted to
the data is the colour temperature, which is approximately equal to the
effective temperature $T$ multiplied by a spectral hardening factor $\xi$.
Values of $\xi \sim 1.7 - 2.6$ have been suggested for Galactic black hole XRBs
\citep[e.g.][]{MerlFabRos00,Makishima00,ShradTit03}.
It has been suggested \citep*{FabRosMill04}
a  spectral correction factor $\xi \approx 1.7$ may be appropriate
for the thermal disk fits to ULX spectra.

\begin{figure}
\centerline{\includegraphics[width=9.0truecm]{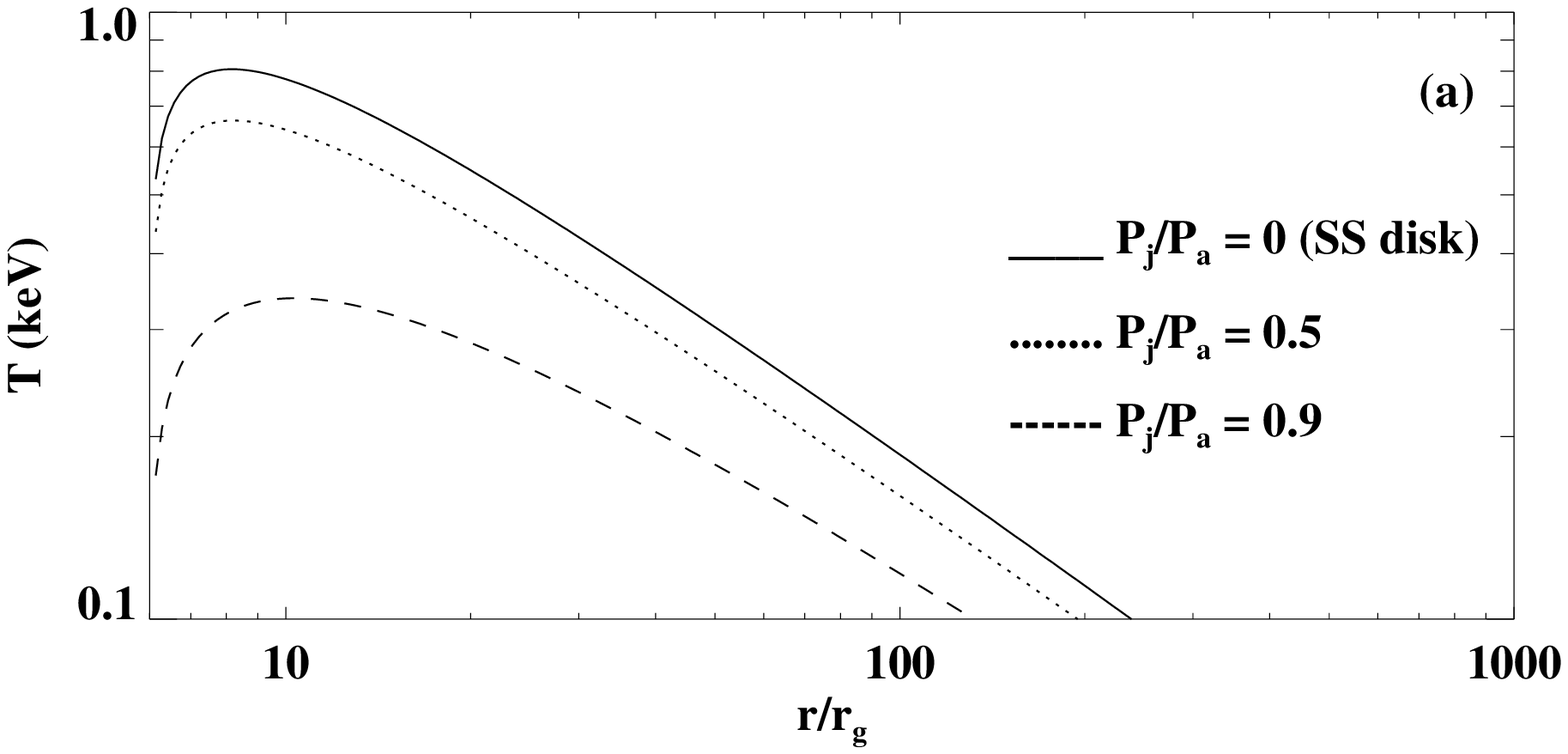}}
\centerline{\includegraphics[width=9.0truecm]{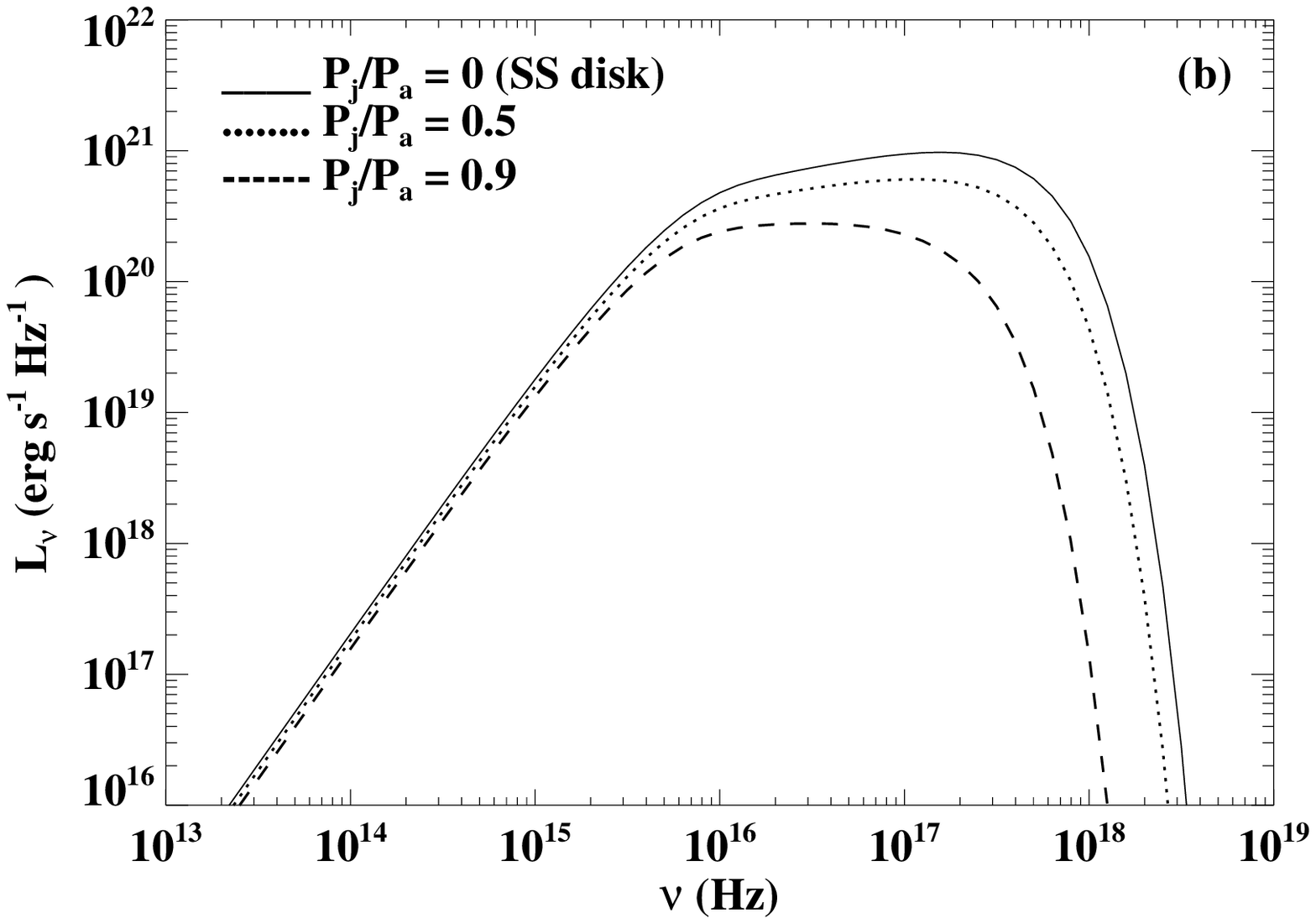}}
\caption{(a) Radial profile of the effective disk temperature, $T(r)$, for a standard
Shakura-Sunyaev (SS) accretion disk (solid line),
and for an outflow--modified disk with $\Pj / \Pa$ values of 0.5 (dotted) and 0.9 (dashed).
(b) Luminosity spectra, $L_{\rm d,\nu}$, for the standard SS multi-colour disk (MCD) and
for the outflow-modified multi-colour disk (OMMCD). Linestyles are the same as in (a).
The black hole mass is $\Mbh = 5M_\odot$, the mass accretion rate is
$\Mdota \simeq 8 \times 10^{-7} \, M_\odot \,{\rm yr}^{-1}$ and the ratio of
outer-to-inner disk radii is $r_{\rm o}/\ri =10^3$.}
\label{f:Tr}
\end{figure}

Figure~\ref{f:Tr} shows plots of $T(r)$ and $L_{\rm d,\nu}$ predicted by our OMMCD model
for $\Mbh = 5M_\odot$, $q=2.6$, $\Pa/\LEdd =1$ and increasing values of $\Pj / \Pa$.
We have integrated to an outer disk radius $r_{\rm o} = 10^3 \ri$.
The mass accretion rate is
$\Mdota = (8\pi m_{\rm p} c \ri /\sigmaT) (\Pa/\LEdd)
\simeq 8 \times 10^{-7} (\ri/6\rg) (\Mbh/5M_\odot) (\Pa/\LEdd)
\, M_\odot \, {\rm yr}^{-1}$.
The value $q=2.6$ is used to demonstrate how the change in radial flux profile affects
the integrated spectrum.
The OMMCD model predicts a decrease in the temperature profile at small radii,
compared with the standard SS disk model (Fig.~1, top panel).
Therefore, the peak in the integrated spectrum comes from larger radii and occurs
at lower energies, for a given $\Mdota$, as $\Pj / \Pa$ increases.
For a given accretion rate and black hole mass, the OMMCD spectrum appears softer and
cooler than a standard disk, and its spectral slope can be flatter than the
characteristic $\nu^{1/3}$ slope of a standard disk spectrum (Fig.~1, bottom panel).
Physically, this occurs because the magnetic disk-jet coupling mechanism is strongest
at small $r$; in the inner disk, more energy is removed from the disk and used to
power a jet.
Phenomenologically, this effect could be modelled in terms of a spectral hardening
factor $\xi \ltapprox 1$.
It has the opposite effect as the hardening factor $\xi \sim 1.7 - 2.6$ sometimes
introduced to account for Comptonization of disk emission \citep{Makishima00,ShradTit03}.

Observationally, the OMMCD model  predicts an anticorrelation between the maximum
disk temperature, $T_{\rm in}$, and the contribution of the hard power-law component,
for a fixed $\LX$, $\Mbh$ and $\Mdota$.
Therefore, it provides an alternative mechanism for producing cooler, fainter disks,
having a somewhat similar effect as higher black hole masses and lower accretion
rates.
This may be relevant for the spectral fitting of some ULXs where an unusually soft X-ray
component ($T_{\rm in} \ltapprox 0.2\,$keV) has been interpreted as disk emission from an
IMBH (see e.g. \citealt*{Miller03,MillFabMill04}; but see also
\citealt{KingPounds03,Crummy05} and \citealt{GonSor06}
for alternative interpretations of this soft component).
We defer a more detailed investigation of the implications of our OMMCD model for ULX
mass estimates to future work (Soria \& Kuncic 2006, in preparation).

More generally, taking into account disk cooling in the presence of a jet can be
important for correct spectral fitting of ULXs as well as Galactic XRBs.
We emphasize that in the OMMCD model, the inner disk is not truncated or replaced
by a radiatively-inefficient flow; therefore, the gravitational energy of  the inflow
is not lost via advection but is mostly transferred to the jet and then partly radiated
with a nonthermal spectrum in the \textit{Chandra} or \textit{XMM-Newton} energy band.
Drainage of the disk power may explain why most ULXs seem to be dominated
by a relatively hard power-law component in the X-ray band even at luminosities
$\gtapprox 0.1 L_{\rm Edd}$; in fact, none of the brightest
ULXs have ever been  observed in a disk-dominated spectral state analagous to the 
high/soft state of Galactic XRBs (see \citealt{Stobbart06} for a more detailed
discussion of this issue).
We note that disk irradiation can be non-negligible when the jet emission is only
weakly beamed \citep{MarkNow04} and this may in part explain the range of $T_{\rm in}$
fits to observed ULX spectra.

\subsection{Jet emission}

A fraction of the total power $\Pj$ channelled from the accretion flow in the form of
Poynting energy is converted into kinetic energy.
This is the basis for all disk-driven jet models \citep[e.g.][]{BlandPayne82}.
Thus, $\Pj$ can be partitioned into bulk and random  particle energies, as well
as magnetic energy \citep[e.g.][]{Celotti98,GhisCelott01}:
\begin{equation}
\Pj = \Pbke + \Pe + \PB
\end{equation}
Some of the kinetic energy can also be subsequently dissipated in the form of radiation.
However, for ultra-relativistic jets (in blazars, for instance) the total radiative output
is usually negligible \citep[see e.g.][]{CelottFab93}.
The kinetic and magnetic energy components are then comoving with approximately constant
speed $\betaj$ and bulk Lorentz factor $\Gammaj = (1 - \betaj^2)^{-1}$.
We consider a conical, quasi-static jet with opening semi-angle $\phij$ and
cross-sectional area $\pi \rj^2$, where $\rj = z \phij$ is the radius at a distance
$z$ along the jet axis.

The jet plasma is quasi-neutral, with one (cold) proton for each (relativistic) electron.
The electrons are assumed to be continuously reaccelerated, maintaining a nonthermal
power-law distribution $\Ne (\gamma) = \Ke \gamma^{-p}$ extending over
energies $\gammamin \ltapprox \gamma \ltapprox \gammamax$ with an average value
$\langle \gamma \rangle = \int \Ne (\gamma) \gamma \, {\rm d}\gamma / \Ne$,
where $\Ne$ is the total electron number density.
A quasi-steady jet requires a stochastic acceleration mechanism, such as second-order
Fermi, in which the accelerating electrons continually scatter off small-scale
inhomogeneities or MHD waves in the jet plasma; in contrast, non-steady jet behaviour
can arise from localized acceleration sites at internal shocks (i.e. via first-order
Fermi -- see Sec. C7 in \citealt*{BBR84} for a detailed discussion on
acceleration in jets).
We use an equipartition relation $\UB = f_{\rm eq} \Ue$ between the magnetic
energy density $\UB = B^2/8\pi$ and electron energy density
$\Ue =  \Ne \langle \gamma \rangle m_{\rm e}c^2$, where the
equipartition factor $f_{\rm eq}$ remains constant along the jet.

The total jet power
is approximately equal to the comoving kinetic and Poynting energy fluxes at the jet
base $z_0$
\citep[e.g.][]{CelottFab93,Celotti98,Schwartz06}:
\begin{eqnarray}
\Pj &\simeq& 2\pi {\rj}_{,0}^2  \Gammaj \betaj c
\left[ (\Gammaj -1){\Ne}_{,0} m_{\rm p}c^2
\right. \nonumber \\
&+&  \left.   
 \frac{4}{3}\Gammaj {\Ne}_{,0} \langle \gamma \rangle m_{\rm e}c^2
+ \Gammaj \frac{B_0^2}{4\pi} \right] 
\label{e:Pj_flux}
\end{eqnarray}
where the subscript `$0$' refers to quantities evaluated at $z_0$.
This relation is used to determine the electron number density at the jet base,
${\Ne}_{,0}$.
The factor of 2 takes into account that $\Pj$ powers two oppositely directed jets.
We use $\phij \simeq 1/\Gammaj$ for $\Gammaj > 1$, consistent with a freely expanding
relativistic flow \citep{BlandKon79}.

The emission spectrum is calculated by subdividing the conical jet into multiple
segments of thickness $\Delta z \ll z$.
There are contributions to the spectrum emitted by each segment from synchrotron radiation,
synchrotron self-Comptonization (SSC) and inverse-Compton (IC) scattering of the disk photons.
We assume the emission to be isotropic in the jet rest-frame. We utilise a simple radiative
transfer prescription to account for synchrotron self-absorption.
The self-Compton and inverse-Compton processes are assumed to be optically thin.
We calculate the (single scattering) IC and SSC emissivities using the full
Klein-Nishina cross-section
\citep[see][]{BlumGould70,Konigl81,Ghisellini85}, although the decline is negligible
in the $0.5 - 8\,$keV bandpass of interest here.
The optically-thin spectral index is
$\alpha \equiv {\rm d}\log L_\nu/{\rm d}\log \nu = -\frac{1}{2}(p-1)$.

\begin{table*}
\centering
\begin{tabular}{|l|c|c|c|c|l|c|c|c|c|}
\hline \\
Model & $\Mbh /M_\odot$ & $\LEdd$ & $\Pa/\LEdd$ & $\Mdota$ &  $\Gammaj$ & $\thetai$ & $\delta$
& $\Lx$ & $\nu L_\nu (5\,{\rm GHz})$ \\
      &           & $({\rm erg \, s}^{-1})$ & &  $(M_\odot \, {\rm yr}^{-1})$  &   & (deg)        
& & (${\rm erg \, s}^{-1}$) & (${\rm erg \, s}^{-1}$) \\
\hline \\
microblazar   & 5 & $6 \times 10^{38}$ & 1.0 & $1 \times 10^{-7}$ &  5   & 5
& 8.4 &  $3 \times 10^{39}$ &  $5 \times 10^{31}$ \\
microquasar         & 20 & $3 \times 10^{39}$ & 1.0 & $5 \times 10^{-7}$ & 4   & 30  
& 1.6 & $5 \times 10^{39}$ &  $8 \times 10^{31}$ \\ 
\hline 
\end{tabular}
\caption{Nominal model parameters resulting in apparent X-ray luminosities 
consistent with a ULX (see text for parameter definitions).
}
\label{t:params}
\end{table*}

We use two different spectral models, depending on whether the jet is viewed
at small or large inclination angles.
For the microblazar model, the inclination of the line of sight to the jet axis is
less than the jet cone-angle (i.e. $\thetai < \phij$).
In this case, the specific luminosity seen by an observer at rest is calculated
by integrating the emissivities over the whole jet length $z_{\rm max}$:
\begin{eqnarray}
\label{e:Lj1}
L_{\rm j,\nu_{\rm obs}}^{\rm obs} &\simeq& 
\int_{z_0}^{z_{\rm max}}   4\pi \delta^2 \left\{ \, 
\epsIC_{\nu = \nu_{\rm obs}/\delta} + \epsSSC_{\nu = \nu_{\rm obs}/\delta}  
\right.  \\
&+& \left.
\epssyn_{\nu = \nu_{\rm obs}/\delta} \exp \left[ 
- \tau_{\nu_{\rm obs}}^{\rm obs}(z) \right]
\, \right\} \pi \rj^2 \, {\rm d}z
\nonumber
\end{eqnarray}
where $\tau_{\nu_{\rm obs}}^{\rm obs} (z) =
\delta^{-1} \kappa_{\nu = \nu_{\rm obs}/\delta}^{\rm syn}(z_{\rm max} - z)$ is the optical
depth to synchrotron self-absorption and
$\delta = [ \Gammaj (1 - \betaj \cos \thetai ) ]^{-1}$
is the relativisitc Doppler factor
\citep[see e.g.][]{Konigl81,LindBland85,Ghisellini93}.
For the microquasar model, we have $\thetai > \phij$.
In this case, the path length through a jet segment as seen by a comoving
observer is $\Delta z / \cos \thetai \simeq \rj (z) /\sin \thetai$.
Assuming $\Delta z$ is sufficiently small that the emissivities
and synchrotron source function $S_\nu^{\rm syn} = \epssyn_\nu / \kappa_\nu^{\rm syn}$
remain approximately constant along each path length, the observed specific
luminosity is then
\begin{eqnarray}
\label{e:Lj2}
L_{\rm j,\nu_{\rm obs}}^{\rm obs} &\simeq&
\sum_{z = z_0}^{z_{\rm max}} 4\pi \left\{
\delta^2 \left[
\epsIC_{\nu = \nu_{\rm obs}/\delta} + \epsSSC_{\nu = \nu_{\rm obs}/\delta} \right]
\frac{\rj}{\sin \thetai} \right. \\
&+& \left.
\delta^3 S_{\nu = \nu_{\rm obs}/\delta}^{\rm syn}
\left[ 1 - \exp \left( - \tau_{\nu_{\rm obs}}^{\rm obs} \right) 
\right]  \right\} \pi \rj^2  \nonumber
\end{eqnarray}
where  the synchrotron self-absorption optical depth is now
$\tau_{\nu_{\rm obs}}^{\rm obs} = \delta^{-1} \kappa_{\nu = \nu_{\rm obs}/\delta}^{\rm syn}
\rj/ \sin \thetai$.

We consider a nonuniform jet in which $\Ne$ decreases with height as
$\Ne (z) \propto z^{-2}$ and hence, $B(z) \propto z^{-1}$ (since $\feq$ is constant),
consistent with flux conservation.
The nonuniformity in $\Ne$ and $B$ means that different jet regions contribute to
different parts of the broadband spectrum, with X-rays produced near the base
and radio emission produced near the opposite end of the jet.
Theoretical spectra predicted by this jet model are presented in the
following section.

\section{Results}
\label{s:results}

The input parameters in our model are:
$\Mbh$, $\Mdota$ (parametrised by $\Pa/\LEdd$), $\Pj/\Pa$ (which normalizes the
magnetic coupling strength, c.f. eqns.~\ref{e:fj} and \ref{e:fjr}), $\Gammaj$, $p$,
$\gammamin$, $\gammamax$, $\feq$, $z_0$, $z_{\rm max}$ and $\thetai$.
Some of these free parameters can be constrained by the observed X-ray and radio
properties of ULXs.
Firstly, we note that soft thermal components, when present in ULX spectra,
contribute no more than $10- 20\,$\% to the total flux \citep[e.g.][]{Stobbart06},
so we set $\Pj / \Pa =0.9$.
Secondly, in the \textit{Chandra} sample \citep{Swartz04}, the average $0.5 - 8\,$keV
spectral index is $\alpha \simeq -0.7$ , implying $p \simeq 2.4$.
Note, however, that the distribution of $\alpha$ in this sample appears to be
approximately Gaussian and extends down to values of $\alpha \simeq 0.0$.

To investigate the radio properties of ULXs, we note that the specific luminosity
of a radio source with flux density $S_\nu$ at a distance $D$ is
$L_\nu \simeq 1.2 \times 10^{24} \left( S_\nu/{\rm mJy} \right)
\left( D/{\rm Mpc} \right)^2 {\rm erg \, s^{-1} \, Hz^{-1}}$.
Radio sources found to be spatially associated with ULXs all have
radio spectral powers well above $10^{24}{\rm erg \, s^{-1} \, Hz^{-1}}$,
while for other nearby ULXs studied by \citet{KordColFalck05}, no radio
counterparts have been found down to a detection limit of
$10^{24}{\rm erg \, s^{-1} \, Hz^{-1}}$.
We shall therefore take $L_\nu = 10^{24}{\rm erg \, s^{-1} \, Hz^{-1}}$ as an
empirical threshold and determine whether either of the micro-blazar/quasar models
can produce radio emission above this level.
If the radio jet emission predicted by these models does not exceed the threshold
for plausible parameters, then the observed radio sources must have other
interpretations (e.g. radio lobes, SNR). 

We consider cases with an equipartition factor $\feq \simeq 1$ and a minimum
electron Lorentz factor $\gammamin = 1$.
The total synchrotron luminosity is most sensitive to these two parameters
and our conservative values are chosen so as to ensure that $L^{\rm syn} \ll P_{\rm jet}$.
The parameter $\gammamax$ determines the high-energy spectral cutoff.
There are no observational constraints on this parameter so we simply
set $\gammamax =10^4$ in all cases.
The jet length, $z_{\rm max}$, determines the synchrotron self-absorption turnover
in the radio and X-ray bands.
We keep this parameter at a fixed value of $10^9 \rg$ so that it scales with $\Mbh$
and gives a radio self-absorption turnover at $\simeq 5$\,GHz.

We differentiate between the microblazar and microquasar scenarios by using the
spectral models defined in Sec. 2 with different  values of the remaining free
parameters: $\Mbh$, $\Mdota$ (parametrized by $\Pa / L_{\rm Edd}$), $\Gammaj$,
and $\thetai$.
Since a fraction $\Pj / \Pa$ of the total accretion power is channelled into a jet,
values of $\Pa / L_{\rm Edd} > 1$ are allowed in the beaming models we consider.
The value of $\Gammaj$ is somewhat constrained by the observation that persistent
jets in Galactic XRBs appear to have $\Gammaj \ltapprox 2$ \citep{Fender05}.
On the other hand, if ULXs are at the extreme end of the XRB population, 
and if they are indeed small-scale versions of blazars and quasars, then we should
not rule out the possibility that $\Gammaj$ could be higher.
The most conservative parameter values that we find can give ULX luminosities are
listed in Table~\ref{t:params}.
The corresponding predicted broadband spectra for the two models are shown in
Figs.~\ref{f:mblazar} and \ref{f:mquasar}.

\subsection{The microblazar model}

\begin{figure}
\centerline{\includegraphics[width=9.0truecm]{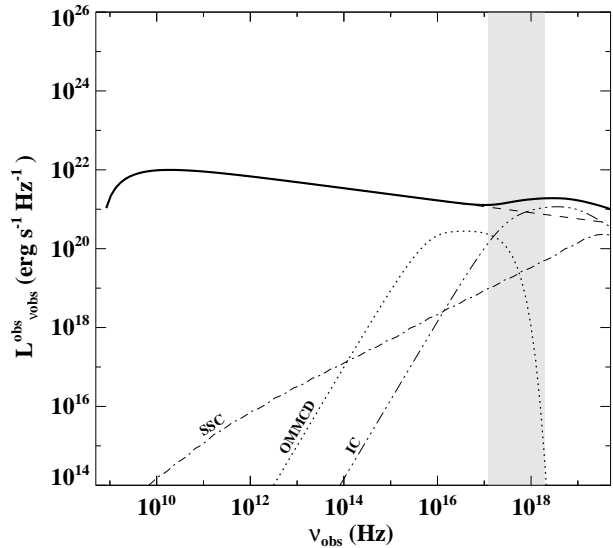}}
\caption{Apparent luminosity spectra predicted for the microblazar model
($\Mbh = 5 M_\odot$, $\thetai < \phij$).
Linestyles are as follows: OMMCD spectrum (dotted), synchrotron (dashed), SSC (dot-dash),
disk IC (dot-dot-dot-dash), and total disk plus jet spectrum (solid). The shaded region
indicates the $0.5 - 8\,$keV X-ray band.}
\label{f:mblazar}
\end{figure}

For this model, we use $\Mbh = 5 M_\odot$,  $\Gammaj = 5$ and $\thetai =5^\circ$,
with the spectral model given by (\ref{e:Lj1}).
This spectral model takes into account self-absorption of synchrotron radiation
along the entire length of the jet.
The resulting X-ray spectrum, Fig.~\ref{f:mblazar}, deviates significantly from a
power-law as a result of relativistically beamed inverse Compton and
(optically-thick) synchrotron emission.
While this is somewhat difficult to reconcile with the generic featureless power-law
fits to most ULX spectra, it is not at all clear whether spectra with $\ltapprox 1000$
counts (e.g. in the \textit{Chandra} sample, \citealt{Swartz04}) have sufficient
signal-to-noise to rule out Comptonization effects altogether
\citep*[see][]{MillFabMill06}.
Indeed, new evidence has emerged from higher signal-to-noise
\textit{XMM-Newton} data for curvature in some ULX spectra above 2\,keV
($\simeq 5 \times 10^{17}\,$)Hz \citep{Stobbart06}.
The spectral curvature has been interpreted in terms of thermal Comptonization from
a warm, optically-thick corona \citep[see e.g.][]{DoneKub05,Goad06}, although this is 
difficult to reconcile with the lack of absorption features in the observed spectra
to date.

The synchrotron optical depth is determined by the total jet length, $z_{\rm max}$.
Here, $z_{\rm max} =10^9 \rg$ corresponds to $\sim 2 \times 10^{-4}$\,pc.
Increasing $z_{\rm max}$ shifts the X-ray self-absorption turnover in Fig.~\ref{f:mblazar}
from $\sim 10^{20}\,$Hz
to even higher frequencies; the highest energy synchrotron photons, produced near
the jet base, have furthest to propagate to reach $z_{\rm max}$ and thus suffer the
most self-absorption.
At the same time, increasing $z_{\rm max}$ shifts the radio turnover to lower frequencies
as more (optically-thin) radio-emitting jet plasma becomes available.
The resulting radio spectrum is approximately flat near 10\,GHz and steepens slightly
towards the mm band.
The opening diameter of the radio-emitting jet is $\sim 8 \mu \,$arcsec at
$D \simeq 1\,$Mpc, consistent with the appearance of a compact core.
Despite the relativistic beaming effects, however, Fig.~\ref{f:mblazar} shows that the
radio spectral power falls short of the estimated $\sim 10^{24} {\rm erg \, s^{-1} \, Hz^{-1}}$
detectability threshold by two orders of magnitude.
Whilst the radio emission can be increased by considering more extreme parameters
(e.g. higher values of $\delta$, $\Pa / \LEdd$, or $\feq$), the resulting X-ray
luminosities become implausibly high and the spectra become too hard.

\subsection{The microquasar model}

One of the difficulties of the microblazar scenario is that it requires an
implausibly large number of sources strongly beamed along different lines of
sight for each source detected as a ULX
(see e.g. the discussion in Sec.~6.2 of \citealt{DavisMush04}).
The microquasar scenario offers a more realistic prospect of addressing the observations
and thus, it is of interest to determine how the emission properties of a jet change as
the inclination angle increases.
In particular, it is not immediately obvious whether a higher $\Mbh$
and $\Mdota$ can completely compensate for the decrease in Doppler boosting to
give ULX X-ray luminosities.

\begin{figure}
\centerline{\includegraphics[width=9.0truecm]{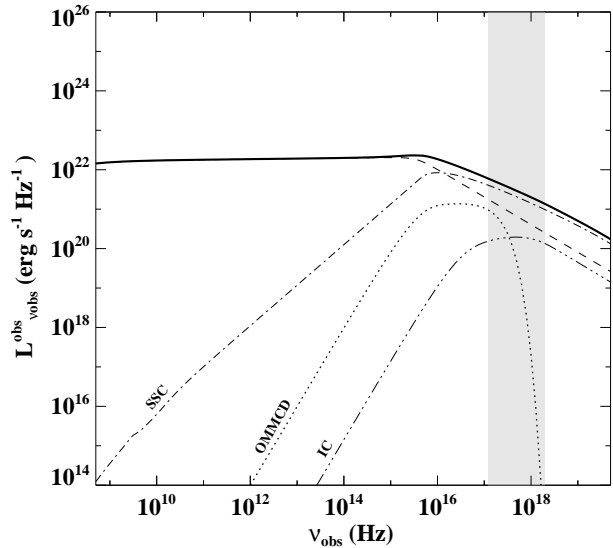}}
\caption{Apparent luminosity spectra predicted for the microquasar model
($\Mbh = 20 M_\odot$, $\thetai > \phij$).
Linestyles are the same as in Fig.~\ref{f:mblazar} and the shaded region
indicates the $0.5 - 8\,$keV X-ray band.
}
\label{f:mquasar}
\end{figure}

We consider  $\Gammaj=4$ with $\thetai =30^\circ$,
giving a jet Doppler factor $\delta = 1.6$.
We also consider a higher mass, $\Mbh = 20 M_\odot$, which is a plausible upper
limit for stellar-mass
black holes formed from individual stars at solar metallicity \citep{FryKal01}.
The spectral model we use for the microquasar case is given by (\ref{e:Lj2})
and the resulting spectrum is shown in Fig.~\ref{f:mquasar}.
The most distinct difference between the microquasar and microblazar models
is that the microquasar model predicts an optically-thin X-ray spectrum.
Additionally, the X-ray spectrum is dominated by SSC emission, rather than
IC emission and the spectrum is slightly flatter than the optically-thin
synchrotron injection spectrum because the SSC photons in the $0.5 - 8\,$keV
band result from seed photons in the optically-thin/thick turnover part of
the synchrotron spectrum ($\sim 10^{15}\,$Hz in Fig.~\ref{f:mquasar}).
As indicated in Table~\ref{t:params}, the predicted X-ray luminosity for the
microquasar model is comparable to that for the microblazar model for the
parameters considered here.

The predicted radio spectral power is also approximately the same.
The jet length, $z_{\rm max} = 10^9 \rg$, corresponds to a physical size $\sim 10^{-3}$\,pc.
For a source at $D \gtapprox 1 \,$Mpc, the projected size of the radio-emitting jet is thus
$\ltapprox 0.2 \,$mas.
While this falls within reach of VLBI, the predicted radio emission of a
microquasar ULX is simply too faint to expect a radio detection, let alone
resolvable radio jet structure.
The predicted specific luminosity at 1\,GHz is
$\sim 10^{22}{\rm erg \, s^{-1} \, Hz^{-1}}$ (comparable to that for the microblazar model),
which corresponds to a flux density $\sim 10 \mu$Jy at 1\,Mpc.
This is well below the detectability threshold and thus, the microquasar model is
consistent with the majority of ULXs not being detected as radio sources.
It also implies that the radio emission found associated with a few ULXs to date
cannot be due to beamed synchrotron emission from a relativistic jet. 

\section{Discussion}
\label{s:discussion}

\subsection{Disk properties}

Our modified disk+jet model predicts \textit{a priori} an anticorrelation between
hard X-ray and soft X-ray spectral components.
In our model, the hard (jet) component is beamed while the soft (disk)
component is isotropic.
Most ULXs appear to be dominated by a single power-law component in the X-ray band
\citep{Swartz04}.
In many cases, the signal-to-noise ratio is not high enough for multi-component
fitting, but even in sources with a high count rate, additional components
(often attributed to thermal disk emission)
contribute typically less than $10- 20\,$\% to the observed flux
\citep{RobCol03,Miller03,MillFabMill04,Krauss05,Stobbart06}.

According to the currently prevailing interpretation of spectral states 
in Galactic XRBs \citep[see][for a review]{McClinRem06},
the low/hard state is accompanied by a steady jet and disk truncation
at radii much larger than the innermost stable circular orbit (ISCO);
this state occurs at low accretion rates ($\Mdota \ltapprox 0.01 \dot{M}_{\rm Edd}$)
and corresponding luminosities $\ltapprox 0.01 L_{\rm Edd}$.
It also implies a lower radiative efficiency in the inner region, often modelled
as an advection-dominated flow.
If ULXs are analogous to Galactic XRBs in their low/hard state
\citep[e.g.][]{Winter06}, then extrapolation of their X-ray luminosities implies
black hole masses $\gtapprox \mbox{a few} \times 10^3 M_\odot$;
it is still not known how such IMBHs can form and whether they are consistent
with the observed ULX population.

We have instead suggested that ULXs behave more as if they are in a ``high/hard''
state, with $\Pa \sim L_{\rm Edd}$, even though $\Ld \ll \Pa$.
In other words,  with a  high accretion rate and gravitational energy released
all the way down to the ISCO, where it is mostly transferred to a steady jet or
corona rather than being locally dissipated and radiated by the disk.
This is different from the high/soft state, in which the jet appears to be
suppressed.
It also differs from the very high (or steep power-law) state that
sometimes accompanies the transition in XRBs from the low/hard to high/soft state
\citep{FendBellGall04,McClinRem06};
the very high state is characterized by sporadic radio ejections and an X-ray
power-law spectrum steeper than those generally measured in bright ULXs.
More generally,
our OMMCD model offers a much needed alternative to the truncated disk model for an
increasing number of sources characterized by a distinctive hard power-law component,
together with evidence  that the accretion flow extends all the way down to the
ISCO \citep{BallFab05,Miller06}.

We predict that, at the typical signal-to-noise level available for most X-ray
spectra of ULXs (generally $\ltapprox 1000$ counts even for the brightest sources),
the modified disk spectral component may or may not be detectable
depending on the relative importance of the jet (the parameter $\Pj/\Pa$ in our
model) and that a low disk temperature may be the result of ``jet cooling'' rather
than an implausibly high black hole mass.
In the calculations presented here, we set  $\Pj / \Pa =0.9$ to produce a
dominant nonthermal hard X-ray spectral component.
To increase the relative contribution of the disk emission to $\LX$,
as appears to be required for some ULXs \citep[e.g.][]{Miller03,Miller04}
we can decrease the ratio $\Pj / \Pa$ of accretion power being diverted into the jet.
The observable disk quantity $T_{\rm in}$ (approximately corresponding to the annulus
with the maximum
flux contribution) also depends on the parameter $q$, which
determines the radial profile of the magnetic torque responsible for jet
formation, as defined in (\ref{e:q}).
We have set $q=2.6$ in the calculations here. 
A steeper profile (e.g. $q \gtapprox 3$) removes more energy from small $r$
and less energy from large $r$, thereby decreasing the X-ray brightness of the disk
for a fixed $\Mdota$.
Thus, to produce ULX spectra with a soft, thermal component
in addition to the hard nonthermal component, as is sometimes observed,
we can decrease $\Pj / \Pa$ in a suitable way to reproduce the observed
fraction of radiative power emitted by the disk, and then vary $q$ to
increase or decrease $T_{\rm in}$.
In general, the radial profile of the magnetic torque does not have to be a scale-free
power-law; we intend to explore different profiles in future work.
For practical purposes, the OMMCD spectrum can also be approximated by a standard disk
with an empirical hardening factor $\xi < 1$ applied to the fitted colour temperature.

It is interesting to consider the implications of our model for the interpretation
of spectral states in Galactic black hole XRBs.
There is increasing evidence from detailed monitoring of individual sources
(e.g. XTE\,J1550$-$564, \citealt{Homan01}; Cyg\,X-1, \citealt{CadBel06};
GX\,339$-$4 \citealt{Miller06})
that changes in $\Mdota$ cannot solely be responsible for spectral state transitions
and that an additional parameter is required to explain the full scope of
observed spectral behaviour in XRBs.
The additional parameter in our model is $\Pj / \Pa$, which measures
the degree of magnetic coupling between disk and jet/corona.
We expect that $\Pj / \Pa$ is approximately independent of $\Mdota$ for the
following reason.
From eqns.~(\ref{e:fj}), (\ref{e:Pa}) and (\ref{e:Pj}), we have
$\Pj / \Pa \propto \ri B_{\phi}^{+}(\ri) B_{z}^{+}(\ri)/(\Mdota \Omega (\ri))
\propto M^2 \Mdota^{-1} B_{\phi}^{+}(\ri) B_{z}^{+}(\ri)$,
since $\ri \propto \rg$.
The magnetic field in the disk is amplified by the fluid shear and becomes dynamically
important when $B^2 \sim \rho v^2$ \citep{KunBick04}.
Since $\rho \propto \Mdota$, we expect 
$\Pj / \Pa \propto M^p$, where $p=2-3$, depending on the details of the accretion
flow velocity and field amplification.
Thus, to first-order, $\Pj / \Pa$ is independent of $\Mdota$ and is an increasing
function of $\Mbh$.
This implies that for a given $\Mdota$, more massive accretors can extract a larger
fraction of total accretion power from the bulk fluid flow to produce more powerful
jets.

Our model thus predicts that, contrary to the standard model for XRBs, steady jets
need not necessarily be quenched in a high state and that sources such as
ULXs which appear to be in a high \textit{and} hard state are likely to have masses
at the high-end of the XRB mass scale.
This prediction is also supported by observations of AGN with
$\Mdota \sim \dot{M}_{\rm Edd}$:
radio-loud quasars tend to have black hole masses at the high end of the supermassive
black hole mass scale (i.e. $\Mbh \sim 10^{8-9}M_\odot$)
\citep[e.g.][]{Laor00,McLureJarvis04}, while narrow-line
Seyfert\,1s (with $\Mbh \sim 10^{6-7} M_\odot$) are preferentially radio-quiet
\citep*[see][and references therein]{GreeHoUlv06,Komossa06}.

Finally, it is worth reminding the reader here that we have worked in a Newtonian potential.
Relativistic effects will change the radial disk flux profile at small $r$ and may restrict
the choice of $q$ and $\Pj/\Pa$.
This may provide further theoretical constraints on the ULX models considered here.
A full relativistic derivation of the OMMCD model is beyond the scope of this paper,
but will be presented elsewhere.

\subsection{Jet properties}

Our results show that beamed emission from a relativistic jet in a stellar-mass black
hole binary system can produce an exceptionally bright X-ray source with a 
$0.5 - 8\,$keV luminosity that is consistent with ULXs.
The predicted X-ray spectra for the microblazar and microquasar scenarios have
different distinguishing properties.
In particular, the microblazar model (where the jet
is viewed close to our line of sight) predicts a spectrum that  becomes increasingly
flatter towards hard X-ray energies in the $0.5 - 8\,$keV bandpass, due to
Comptonization effects (see Fig.~\ref{f:mblazar}).
The observed spectrum of once-scattered disk photons rises to a peak
just beyond the \textit{Chandra}/\textit{XMM-Newton} bandpass.
The power-law component of this beamed inverse Compton emission falls in the gamma-ray band.
It is difficult to rule out or confirm possible Comptonization effects from currently available
ULX spectra (\citealt{MillFabMill06}; see also the discussion in \citealt{GonSor06}).
We do note, however, that some evidence for spectral curvature towards hard X-ray
energies has recently emerged in \textit{XMM-Newton} ULX spectra with high photon counts
\citep{Stobbart06}.
The microquasar model, on the other hand, predicts a power-law X-ray spectrum due to
beamed synchrotron self-Comptonization and
thus offers a viable alternative to thermal Comptonization in a corona.

The predicted  radio properties of the micro-blazar/quasar models are similar:
the spectra are approximately flat near $5- 10 \,$GHz and the specific radio luminosities
are  $\sim 10^{22} \, {\rm erg \, s^{-1} \, Hz^{-1}}$, corresponding to
$\sim 10 \mu \,$Jy at $D \sim 1 \,$Mpc.
How do these predicted radio properties compare to those observed?
Very few radio counterparts of ULXs have been found so far, down to
detection limits $\sim 10^{24} \, {\rm erg \, s^{-1} \, Hz^{-1}}$.
The very low detection rates resulting from radio counterpart searches
\citep[e.g.][]{Ghosh05,KordColFalck05} strongly suggest that the vast majority
of ULXs are intrinsically weak radio sources.
Indeed, the ratio of radio-to-X-ray powers,
$\RX \equiv \nu L_\nu (5\,{\rm GHz})/\LX$  \citep{TeraWil03},
predicted by our spectral modelling is $\RX \simeq 10^{-8}$
for  both the microblazar and microquasar cases (see Table~\ref{t:params}).
This is three orders of magnitude below the peak radio-to-X-ray
power ratio observed in Galactic XRBs \citep{FendKuul01}.
Note, however, this also implies that a low measured value of $\RX$ for
ULXs with a radio counterpart cannot be used to rule out relativistic beaming
\citep[see e.g.][]{MillFabMill04}.


Our results imply that if the radio emission arises from synchrotron radiation in a
relativistic jet, then ULXs should be too faint to be detected by current
radio surveys.
The few radio sources found associated with ULXs to date have specific radio powers
$\sim 10^{25} \, {\rm erg \, s^{-1} \, Hz^{-1}}$ and steep radio spectra:
$\alpha \sim -1$ for the ULX in NGC~5408 \citep{Kaaret03,Soria06a},
$\alpha \sim -(0.4-0.5)$ for the ULX in Ho~II \citep{TonWest95,MillMushNeff05}
and $\alpha \sim -0.65$ for ULX\,2 in NGC~7424 \citep{Soria06b}.
There is also evidence that the observed radio emission is marginally
resolved (extended over $\sim 30-50\,$pc) in two of those sources (NGC~5408 and Ho~II)
and for all the radio counterparts found so far, the ULX lies in an active
star-forming region.
The high flux, steep spectral index and possible spatial resolution suggest that
those detected radio sources are not directly associated with a relativistic jet, but
instead probably result from optically-thin emission from  radio lobes or a radio SNR
located near the ULX \citep{MillMushNeff05,Soria06a,Soria06b}
or even a combination of both (e.g. a scaled-up version of the Galactic source
SS433; \citealt{Fabrika04}).

In the case of the radio lobe interpretation, the low detection rate can be
attributed to the local ISM environment, which may be sufficiently less dense than
the IGM that adiabatic losses are favoured over synchrotron losses in the jet lobes.
This may explain the paucity of radio lobes in Galactic XRBs relative to those
commonly seen in radio galaxies and quasars  \citep{Heinz02,Hardcastle05}.

Finally, it is interesting to note that the theoretical radio and X-ray luminosities
predicted by our relativistic beaming model for ULXs are remarkably
consistent with the ``fundamental plane'' correlation
between radio and X-ray emission and black hole mass
\citep*{GallFendPool03,MerlHeindiMat03,FalckKordMark04}.
This is surprising for two reasons.
Firstly, the theoretical relation derived by \citet{FalckKordMark04} assumes
optically-thin X-ray emission and neglects differences in radio and X-ray beaming
(i.e. equivalent to our microquasar spectral model).
It is not immediately obvious why the relation should hold if part of the X-ray
emission is optically-thick, as is the case for our microblazar model.
Secondly, the empirical relation found by \citet{MerlHeindiMat03},
\textit{viz.} $\log \LR \simeq 0.60 \log \LX + 0.78 \log \Mbh + 7.33$
(where $\LR$ denotes radio power at the observing frequency, $\nu L_\nu$),
strictly only holds for sources in a low state;
it is still unclear to what extent it holds or breaks down for sources 
which are in a high state \citep*[see][]{KordFalckCorb06}.

\section{Conclusions}
\label{s:conclusions}

Current observational evidence suggests that most of the X-ray emission from ULXs does not
come directly from the accretion disk, which is either not directly visible, or colder
than in bright stellar-mass XRBs.
We have used a coupled disk+jet theoretical framework to explain why most of the energy
is not directly radiated from the disk, and to test whether a relativistic beaming
scenario (microblazar or microquasar) is consistent with the observed X-ray and radio
spectra.
Our main conclusions can be summarized as follows:
\begin{list}{}{\itemsep=0pt}
\item[1.] The accretion disk can be substantially modified by the presence of a
magnetized jet; this gives rise not only to a modified disk spectrum,
but also to a decrease in the peak colour temperature ($T_{\rm in}$) of the inner disk
for a fixed mass accretion rate.
The modified disk+jet model thus predicts an anticorrelation between the relative
importance of hard (jet) and soft (disk) X-ray spectral components.
\item[2.] We confirm that both the stellar-mass microblazar and microquasar scenarios can
produce apparent X-ray luminosities in the ULX regime
($\Lx \gtapprox 2 \times 10^{39} \, {\rm erg \, s^{-1}}$).
A general feature of our disk+jet model is that it can produce hard, nonthermal
X-ray spectra that are analogous to the classic low/hard spectral state in XRBs,
but applicable to high mass accretion rates ($\Mdota \sim \dot M_{\rm Edd}$) and,
therefore, high luminosity sources.
In our model, a magnetic torque provides the mechanism for transferring a substantial fraction
of the total accretion power $\Pa \sim L_{\rm Edd}$ from the disk to the jet
(i.e. $\Pj \sim L_{\rm Edd}$).
Our spectral model for the microblazar predicts a substantially hardened X-ray spectrum
due to strong beaming and Comptonization effects.
The microquasar model, on the other hand, produces a featureless power-law spectrum.
\item[3.] Both the microblazar and microquasar models predict that,
despite the beaming effects, ULXs are intrinsically weak radio emitters.
This result is compatible with the overwhelming excess of non-detections
over detections in ULX radio observations to date.
Neither of the beaming models are consistent with the few ULX radio counterpart
detections reported to date.
We therefore conclude that the observed radio emission cannot be attributed to
beamed synchrotron emission in a relativistic jet.
We suggest it is likely due to radio lobes or a radio SNR.
We predict that deeper observations around 20\,GHz should in principle detect
a change in radio spectral index, thus resolving the direct jet contribution
from the lobe emission.
\end{list}

In this paper, we have restricted ourselves to modelling beamed emission from
accreting black holes with masses in the known stellar-mass range
($5-20 \, M_\odot$).
If IMBHs exist, it is legitimate to assume (based on the stellar-mass and supermassive
black hole analogies) that  some IMBHs can also possess relativistic jets.
If some IMBHs have beamed emission, then we would expect to find at least a
few milliblazars and milliquasars with apparent X-ray luminosities in excess of
$10^{41} \, {\rm erg \, s^{-1}}$.
Such sources have not yet been seen.

The disk+jet model we have proposed here requires that more gravitational energy is
channelled into a jet
than is radiated by a disk, at least during some phases of accretion.
At low or moderate accretion rates, new evidence of significant feedback
onto the ISM surrounding accreting black holes indicates that jets can indeed
carry away much more energy than is radiated by the disk
\citep[e.g.][]{Owen00,Churazov02}.
At accretion rates approaching the Eddington rate, however, observations
suggest that jets are quenched in Galactic XRBs \citep[e.g.][]{FendBellGall04}
and in low-mass AGN \citep[e.g.][]{GreeHoUlv06},
but can persist in AGN with high black hole masses
\citep[e.g.][]{Laor00,McLureJarvis04}.
Our model predicts that the fractional accretion power chanelled into jets scales
with black hole mass and is independent of the mass accretion rate.
It therefore predicts that jets can be the dominant energy carrier, even at high
($\sim$ Eddington) accretion rates.
If this is correct, it should be possible to find evidence of relativistic jet emission
also in some other accreting black hole sources in which the hard X-ray component
has perhaps been misinterpreted as, for example, thermal Comptonization in a corona.

\section*{Acknowledgments}

ZK thanks K. Wu, R. Hunstead and E. Sadler for useful discussions and acknowledges
support from a University of Sydney Research and Development Grant.
RS acknowledges support from a University of Sydney Denison Grant and an OIF Marie
Curie Fellowship.
GVB acknowledges support for this research from Australian Research Council
Discovery Project 0345983.

\bsp
\label{lastpage}

\end{document}